\documentclass{article}

\usepackage[dvips]{graphicx}
\usepackage{cite}
\usepackage{hyperref}

\title{Calculations of widths in the problem of decay by proton emission}

\author{Sergei~P.~Maydanyuk\thanks{\emph{E-mail:} maidan@kinr.kiev.ua}\, and
Sergei~V.~Belchikov\thanks{\emph{E-mail:} sbelchik@kinr.kiev.ua} \\
\small{\emph{Institute for Nuclear Research, National Academy of Science of Ukraine, Kiev, 03680, Ukraine}}}

\date{\small\today}

\begin{document}

\maketitle

\begin{abstract}
We develop a new fully quantum method for determination of widths for nuclear decay by proton emission where multiple internal reflections of wave packet describing tunneling process inside proton--nucleus radial barrier are taken into account.
Exact solutions for amplitudes of wave function, penetrability $T$ and reflection $R$ are found for $n$-step barrier (at arbitrary $n$) which approximates the realistic barrier.
In contrast to semiclassical approach and two-potential approach, we establish by this method essential dependence of the penetrability on the starting point $R_{\rm form}$ in the internal well where proton starts to move outside (for example, for $^{157}_{73}{\rm Ta}$ the penetrability is changed up to 200 times; accuracy is $|T+R-1| < 1.5 \cdot 10^{-15}$).
We impose a new condition: in the beginning of the proton decay the proton starts to move outside from minimum of the well. Such a condition provides minimal calculated half-life and gives stable basis for predictions. However, the half-lives calculated by such an approach turn out to be a little closer to experimental data in comparison with the semiclassical half-lives.
Estimated influence of the external barrier region is up to 1.5 times for changed penetrability.
\end{abstract}




\section{Introduction
\label{introduction}}

Nuclei beyond the proton drip line are ground-state proton emitters, i.~e. nuclei unstable for emission of proton from the ground state.
Associated lifetimes, ranging from $10^{-6}$~sec to few seconds, are sufficiently long to obtain wealth of spectroscopic information.
Experimentally, a number of proton emitters has been discovered in the mass region $A \approx 110$, 150, and 160 (see \cite{Hofmann.1989,Hofmann.1995.RCA,Hofmann.1996,Davids.1996.PRL} and references in cited papers).
A new regions of proton unstable nuclei is supposed to be explored in close future using radioactive nuclear beams.

Initially, the parent nucleus is in quasistationary state, and the proton decay may be considered as a process where the proton tunnels through potential barrier.
In theoretical study one can select three prevailing approaches \cite{Aberg.1997.PRC}: approach with distorted wave Born approximation (DWBA), two-potential approach (TPA), and approach for description of penetration through the barrier in terms of one-dimensional semiclassical method (WKBA).
In systematical study these approaches are correlated between themselves, while calculation of penetrability of the barrier is keystone in successful estimation of gamma widths.
While the third approach studies such a question directly, in the first and second approaches the penetrability of the barrier is not studied and the width is based on correlation between wave functions in the initial and final states.
However, the most accurate information on amplitudes and phases of these wave functions and correspondence between them can be obtained from unite picture of penetration of proton through the barrier, which is used in the WKBA approach (up to the second order approximation).


The main objective of this paper is to pass from semiclassical unite description of the process of penetration of proton through the barrier used in the WKBA approach to its fully quantum analogue, to put a fully quantum grounds for determination of the penetrability in this problem.
In order to provide such a formalism, we have improved method of multiple internal reflections (MIR, see Refs.~\cite{Maydanyuk.2000.UPJ,Maydanyuk.2002.JPS,Maydanyuk.2002.PAST,Maydanyuk.2003.PhD-thesis,Maydanyuk.2006.FPL}) generalizing it
on the radial barriers of arbitrary shapes.
In order to realize this difficult improvement, we have restricted ourselves by consideration of the spherical ground-state proton emitters, while nuclear deformations are supposed to be further included by standard way.
This advance of the method never studied before allows to describe dynamically a process of penetration of the proton through the barrier of arbitrary shape in fully quantum consideration, to calculate penetrability and reflection without the semiclassical restrictions, to analyze abilities of the semiclassical and other models on such a basis.

This paper is organized in the following way. In Sec.~\ref{sec.2}, formalism of the method of multiple internal reflections in description of tunneling of proton through the barrier in proton decay is presented. Here, we give solutions for amplitudes, define penetrability, width and half-life.
In Sec.~\ref{sec.3}, results of calculations are confronted with experimental data and are compared with semiclassical ones. Here, using the fully quantum basis of the method, we study a role of the barrier shape in calculations of widths in details. In particular, we observe essential influence of the internal well before the barrier on the penetrability. We discuss shortly possible interconnections between the proposed approach and other fully quantum methods of calculation of widths.
In Sec.~\ref{conclusions}, we summarize results.
Appendixes include proof of the method MIR and alternative standard approach of quantum mechanics used as test for the method MIR and for the results presented.


\section{Theoretical approach
\label{sec.2}}


An approach for description of one-dimensional motion of a non-relativistic particle above a barrier on the basis of multiple internal reflections of stationary waves relatively boundaries has been studied in number of papers and is known (see \cite{Fermor.1966.AJPIA,McVoy.1967.RMPHA,Anderson.1989.AJPIA} and references therein).
Tunneling of the particle under the barrier was described successfully on the basis of multiple internal reflections of the wave packets relatively boundaries (approach was called as \emph{method of multiple internal reflections} or \emph{method MIR}, see Refs.~\cite{Maydanyuk.2000.UPJ,Maydanyuk.2002.JPS,Maydanyuk.2002.PAST,Maydanyuk.2003.PhD-thesis}).
In such approach it succeeded in connecting:
1) continuous transition of solutions for packets after each reflection, total packets between the above-barrier motion and the under-barrier tunneling;
2) coincidence of transmitted and reflected amplitudes of stationary wave function in each spatial region obtained by approach MIR with the corresponding amplitudes obtained by standard method of quantum mechanics;
3) all non-stationary fluxes in each step, are non-zero that confirms propagation of packets under the barrier (i.~e. their ``tunneling'').
In frameworks of such a method, non-stationary tunneling obtained own interpretation, allowing to study this process at interesting time moment or space point. In calculation of phase times this method turns out to be enough simple and convenient \cite{Maydanyuk.2006.FPL}. It has been adapted for scattering of the particle on nucleus and $\alpha$-decay in the spherically symmetric approximation with the simplest radial barriers~\cite{Maydanyuk.2000.UPJ,Maydanyuk.2002.JPS,Maydanyuk.2003.PhD-thesis} and for tunneling of photons~\cite{Maydanyuk.2002.JPS,Maydanyuk.2006.FPL}.
However, further realization of the MIR approach meets with three questions.
1) \emph{Question on effectiveness.}
The multiple reflections have been proved for the motion above one rectangular barrier and for tunneling under it~\cite{Anderson.1989.AJPIA,Maydanyuk.2002.JPS,Maydanyuk.2006.FPL}. However, after addition of the second step it becomes unclear how to separate the needed reflected waves from all their variety in calculation of all needed amplitudes. After obtaining exact solutions of the stationary amplitudes for two arbitrary rectangular barriers \cite{Maydanyuk.2003.PhD-thesis,Maydanyuk.2000.UPJ}, it becomes unclear how to generalize such approach for barriers with arbitrary complicate shape. In Ref.~\cite{Esposito.2003.PRE} multiple internal reflections of the waves were studied for tunneling through a number of equal rectangular steps separated on equal distances. However, the amplitudes were presented for two such steps only, in approximation when they were separated on enough large distance, and these solutions in approach of multiple internal reflections were based of the amplitudes of total wave function obtained before by standard method (see Appendix A, eqs.~(7), (18), (19) in this paper). So, \emph{we come to a serious unresolved problem of realization of the approach of multiple reflections in real quantum systems with complicated barriers}, and clear algorithms of calculation of amplitudes should be constructed.

2) \emph{Question on correctness.}
Whether is interference between packets formed relatively different boundaries appeared? Whether does this come to principally different results of the approach of multiple internal reflections and direct methods of quantum mechanics? Note that such interference cannot be appeared in tunneling through one rectangular barrier and, therefore, it could not visible in the previous papers.

3) \emph{Question on uncertainty in radial problem.}
Calculations of half-lives of different types of decays based on the semiclassical approach are prevailing today.
For example, in Ref.~\cite{Buck.1993.ADNDT} agreement between experimental data of $\alpha$-decay half-lives and ones calculated by theory is demonstrated in a wide region of nuclei from $^{106}{\rm Te}$ up to nucleus with $A_{d}=266$ and $Z_{d}=109$ (see Ref.~\cite{Denisov.2005.PHRVA} for some improved approaches). In review~\cite{Sobiczewski.2007.PPNP} methodology of calculation of half-lives for spontaneous-fission is presented (see eqs.~(21)--(24) in p.~321).
Let us consider proton-decay of nucleus where proton penetrates from the internal region outside with its tunneling through the barrier. \underline{At the same boundary condition}, reflected and incident waves turn out to be defined with uncertainty. How to determine them?
The semiclassical approach gives such answer:
\emph{according to theory, in construction of well known formula for probability we neglect completely by the second (increasing) item of the wave function inside tunneling region} (see Ref.~\cite{Landau.v3.1989}, eq.~(50.2), p.~221). In result, equality $T^{2} + R^{2} = 1$ has no any sense (where $T$ and $R$ are coefficients of penetrability and reflection). Condition of continuity for the wave function and for total flux is broken at turning point. So, we do not find reflection $R$. We do not suppose on possible interference between incident and reflected waves which can be non zero. The penetrability is determined by the barrier shape inside tunneling region, while internal and external parts do not take influence on it. The penetrability does not dependent on depth of the internal well (while the simplest rectangular well and barrier give another exact result).
But, the semiclassical approach is so prevailing that one can suppose that it has enough well approximation of the penetrability estimated. It turns out that if in fully quantum approach to determine the penetrability through the barrier (constructed on the basis of realistic potential of interaction between proton and daughter nucleus) then one can obtain answer ``no''. Fully quantum penetrability is a function of new additional independent parameters,
it can achieve essential difference from semiclassical one (at the same boundary condition imposed on the wave function).
This will be demonstrated below.


\subsection{Decay with radial barrier composed from arbitrary number of rectangular steps
\label{sec.2.2}}

Let us assume that starting from some time moment before decay the nucleus could be considered as system composite from daughter nucleus and fragment emitted. It`s decay is described by a particle with reduced mass $m$ which moves in radial direction inside a radial potential with a barrier. We shall be interesting in the radial barrier of arbitrary shape,
which has successfully been approximated by finite number $N$ of rectangular steps:
\begin{equation}
  V(r) = \left\{
  \begin{array}{cll}
    V_{1},   & \mbox{at } R_{\rm min} < r \leq r_{1}      & \mbox{(region 1)}, \\
    V_{2},   & \mbox{at } r_{1} \leq r \leq r_{2}         & \mbox{(region 2)}, \\
    \ldots   & \ldots & \ldots \\
    V_{N},   & \mbox{at } r_{N-1} \leq r \leq R_{\rm max} & \mbox{(region N)},
  \end{array} \right.
\label{eq.2.2.1}
\end{equation}
where $V_{i}$ are constants ($i = 1 \ldots N$). We define the first region 1 starting from point $R_{\rm min}$, assuming that the fragment is formed here and then it moves outside. We shall be interesting in solutions for above barrier energies while the solution for tunneling could be obtained after by change
$i\,\xi_{i} \to k_{i}$. A general solution of the wave function (up to
its normalization) has the following form:
\begin{equation}
  \psi(r, \theta, \varphi) =  \frac{\chi(r)}{r} Y_{lm}(\theta, \varphi),
\label{eq.2.2.2}
\end{equation}
\begin{equation}
\chi(r) = \left\{
\begin{array}{lll}
   e^{ik_{1}r} + A_{R}\,e^{-ik_{1}r}, & \mbox{at } R_{\rm min} < r \leq r_{1} & \mbox{(region 1)}, \\
   \alpha_{2}\, e^{ik_{2}r} + \beta_{2}\, e^{-ik_{2}r}, & \mbox{at } r_{1} \leq r \leq r_{2} & \mbox{(region 2)}, \\
   \ldots & \ldots & \ldots \\
   \alpha_{n-1}\, e^{ik_{N-1}r} + \beta_{N-1}\, e^{-ik_{N-1}r}, &
     \mbox{at } r_{N-2} \leq r \leq r_{N-1} & \mbox{(region N-1)}, \\
   A_{T}\,e^{ik_{N}r}, & \mbox{at } r_{N-1} \leq r \leq R_{\rm max} & \mbox{(region N)},
\end{array} \right.
\label{eq.2.2.3}
\end{equation}
where $\alpha_{j}$ and $\beta_{j}$ are unknown amplitudes, $A_{T}$ and $A_{R}$ are unknown amplitudes of transmission and reflection, $Y_{lm}(\theta, \varphi)$ is spherical function, $k_{i} = \frac{1}{\hbar}\sqrt{2m(E-V_{i})}$ are complex wave vectors. We shall be looking for solution for such problem in approach of multiple internal reflections (we restrict ourselves by a case of orbital moment $l=0$ while its non-zero generalization changes the barrier shape which was used as arbitrary before in development of formalism MIR and, so, is absolutely non principal).

According to the method of multiple internal reflections, scattering of the particle on the barrier is considered on the basis of wave packet consequently by steps of its propagation relatively to each boundary of the barrier (idea of such approach can be understood the most clearly in the problem of tunneling through the simplest rectangular barrier, see \cite{Maydanyuk.2002.JPS,Maydanyuk.2003.PhD-thesis,Maydanyuk.2006.FPL} and Appendix A where one can find proof of the method and analysis of its properties). Each step in such consideration of propagation of the packet will be similar to on from the first $2N-1$ steps, independent between themselves. From analysis of these steps recurrent relations are found for calculation of unknown amplitudes $A^{(n)}$, $S^{(n)}$, $\alpha^{(n)}$ and $\beta^{(n)}$ for arbitrary step $n$, summation of these amplitudes are calculated.
We shall be looking for the unknown amplitudes, requiring wave function and its derivative to be continuous at each boundary. We shall consider the coefficients $T_{1}^{\pm}$, $T_{2}^{\pm}$, $T_{3}^{\pm}$ and $R_{1}^{\pm}$, $R_{2}^{\pm}$, $R_{3}^{\pm}$ as additional factors to amplitudes $e^{\pm i\,k\,x}$. Here, bottom index denotes number of the region, upper (top) signs ``$+$'' and ``$-$'' denote directions of the wave to the right or to the left, correspondingly. At the first, we calculate $T_{1}^{\pm}$, $T_{2}^{\pm}$ \ldots $T_{N-1}^{\pm}$ and $R_{1}^{\pm}$,
$R_{2}^{\pm}$ \ldots $R_{N-1}^{\pm}$:
\begin{equation}
\begin{array}{ll}
\vspace{2mm}
   T_{j}^{+} = \displaystyle\frac{2k_{j}}{k_{j}+k_{j+1}} \,e^{i(k_{j}-k_{j+1}) r_{j}}, &
   T_{j}^{-} = \displaystyle\frac{2k_{j+1}}{k_{j}+k_{j+1}} \,e^{i(k_{j}-k_{j+1}) r_{j}}, \\
   R_{j}^{+} = \displaystyle\frac{k_{j}-k_{j+1}}{k_{j}+k_{j+1}} \,e^{2ik_{j}r_{j}}, &
   R_{j}^{-} = \displaystyle\frac{k_{j+1}-k_{j}}{k_{j}+k_{j+1}} \,e^{-2ik_{j+1}r_{j}}.
\end{array}
\label{eq.2.2.4}
\end{equation}
Using recurrent relations:
\begin{equation}
\begin{array}{l}
   \vspace{1mm}
   \tilde{R}_{j-1}^{+} =
     R_{j-1}^{+} + T_{j-1}^{+} \tilde{R}_{j}^{+} T_{j-1}^{-}
     \Bigl(1 + \sum\limits_{m=1}^{+\infty} (\tilde{R}_{j}^{+}R_{j-1}^{-})^{m} \Bigr) =
     R_{j-1}^{+} +
     \displaystyle\frac{T_{j-1}^{+} \tilde{R}_{j}^{+} T_{j-1}^{-}} {1 - \tilde{R}_{j}^{+} R_{j-1}^{-}}, \\

   \vspace{1mm}
   \tilde{R}_{j+1}^{-} =
     R_{j+1}^{-} + T_{j+1}^{-} \tilde{R}_{j}^{-} T_{j+1}^{+}
     \Bigl(1 + \sum\limits_{m=1}^{+\infty} (R_{j+1}^{+} \tilde{R}_{j}^{-})^{m} \Bigr) =
     R_{j+1}^{-} +
     \displaystyle\frac{T_{j+1}^{-} \tilde{R}_{j}^{-} T_{j+1}^{+}} {1 - R_{j+1}^{+} \tilde{R}_{j}^{-}}, \\

   \tilde{T}_{j+1}^{+} =
     \tilde{T}_{j}^{+} T_{j+1}^{+}
     \Bigl(1 + \sum\limits_{m=1}^{+\infty} (R_{j+1}^{+} \tilde{R}_{j}^{-})^{m} \Bigr) =
     \displaystyle\frac{\tilde{T}_{j}^{+} T_{j+1}^{+}} {1 - R_{j+1}^{+} \tilde{R}_{j}^{-}},
\end{array}
\label{eq.2.2.5}
\end{equation}
and selecting as starting the following values:
\begin{equation}
\begin{array}{ccc}
  \tilde{R}_{N-1}^{+} = R_{N-1}^{+}, &
  \tilde{R}_{1}^{-} = R_{1}^{-}, &
  \tilde{T}_{1}^{+} = T_{1}^{+},
\end{array}
\label{eq.2.2.6}
\end{equation}
we calculate successively coefficients $\tilde{R}_{N-2}^{+}$ \ldots $\tilde{R}_{1}^{+}$, $\tilde{R}_{2}^{-}$ \ldots
$\tilde{R}_{N-1}^{-}$ and $\tilde{T}_{2}^{+}$ \ldots $\tilde{T}_{N-1}^{+}$.
At finishing, we determine coefficients $\beta_{j}$:
\begin{equation}
\begin{array}{l}
   \vspace{1mm}
   \beta_{j} =
     \tilde{T}_{j-1}^{+}
     \Bigl(1 + \sum\limits_{m=1}^{+\infty} (\tilde{R}_{j}^{+} \tilde{R}_{j-1}^{-})^{m} \Bigr) =
     \displaystyle\frac{\tilde{T}_{j-1}^{+}} {1 - \tilde{R}_{j}^{+} \tilde{R}_{j-1}^{-}},
\end{array}
\label{eq.2.2.7}
\end{equation}
the amplitudes of transmission and reflection:
\begin{equation}
\begin{array}{cc}
  A_{T} = \tilde{T}_{N-1}^{+}, &
  A_{R} = \tilde{R}_{1}^{+}
\end{array}
\label{eq.2.2.8}
\end{equation}
and corresponding coefficients of penetrability $T$ and reflection $R$:
\begin{equation}
\begin{array}{cc}
  T_{MIR} = \displaystyle\frac{k_{n}}{k_{1}}\; \bigl|A_{T}\bigr|^{2}, &
  R_{MIR} = \bigl|A_{R}\bigr|^{2}.
\end{array}
\label{eq.2.2.9}
\end{equation}
We check the property:
\begin{equation}
\begin{array}{ccc}
  \displaystyle\frac{k_{n}}{k_{1}}\; |A_{T}|^{2} + |A_{R}|^{2} = 1 & \mbox{ or }&
  T_{MIR} + R_{MIR} = 1.
\end{array}
\label{eq.2.2.10}
\end{equation}
which should be the test, whether the method MIR gives us proper
solution for wave function. Now if energy of the particle is
located below then height of one step with number $m$, then for
description of transition of this particle through such barrier
with its tunneling it shall need to use the following change:
\begin{equation}
  k_{m} \to i\,\xi_{m}.
\label{eq.2.2.11}
\end{equation}
For the potential from two rectangular steps (with different choice of their sizes) after comparison between the all amplitudes obtained by method of MIR and the corresponding amplitudes obtained by standard approach of quantum mechanics, we obtain coincidence up to first 15 digits. Increasing of number of steps up to some thousands keeps such accuracy and fulfillment of the property (\ref{eq.2.2.10}) (see Appendix~\ref{sec.app.2} where we present shortly the standard technique of quantum mechanics applied for the potential (\ref{eq.2.2.1}) and all obtained amplitudes). This is important test which confirms reliability of the method MIR. So, we have obtained full coincidence between all amplitudes, calculated by method MIR and by standard approach of quantum mechanics, and that is way we generalize the method MIR for description of tunneling of the particle through potential, consisting from arbitrary number of rectangular barriers and wells of arbitrary shape.

%

\subsection{Width $\Gamma$ and half-life
\label{sec.2.3}}

We define width $\Gamma$ of the decay of the studied quantum system by following the procedure of Gurvitz and
K\"{a}lbermann \cite{Gurvitz.1987.PRL}:
\begin{equation}
  \Gamma = P_{p}\, F\: \displaystyle\frac{\hbar^{2}}{4m}\; T,
\label{eq.2.3.1}
\end{equation}
where $P_{p}$ is the preformation probability and $F$ is the
normalization factor. $T$ is the penetrability coefficient in
propagation of the particle from the internal region outside with
its tunneling through the barrier, which we shall calculate by
approach MIR or by approach WKB. In approach WKB we define it so:
\begin{equation}
  T_{WKB} =
  \exp\;
  \Biggl\{
    -2 \displaystyle\int\limits_{R_{2}}^{R_{3}}
    \sqrt{\displaystyle\frac{2m}{\hbar^{2}}\: \Bigl(Q - V(r)\Bigr)} \; dr
  \Biggr\}
\label{eq.2.3.2}
\end{equation}
where $R_{2}$ and $R_{3}$ are the second and third turning points.
According to~\cite{Buck.1993.ADNDT}, the \emph{normalization factor} $F$ is given by simplified way by $F_{1}$
or by improved way by $F_{2}$ so:
\begin{equation}
\begin{array}{ll}
  F_{1} =
  \Biggl\{\:
    \displaystyle\int\limits_{R_{1}}^{R_{2}}
    \displaystyle\frac{dr}{2k(r)}
  \Biggr\}^{-1}, &
  F_{2} =
  \Biggl\{\:
    \displaystyle\int\limits_{R_{1}}^{R_{2}}
    \displaystyle\frac{1}{k(r)}\;
    \cos^{2} \Biggl[\:\displaystyle\int\limits_{R_{1}}^{r} k(r')\; dr' - \displaystyle\frac{\pi}{4} \Biggr]\; dr
  \Biggr\}^{-1}.
\end{array}
\label{eq.2.3.3}
\end{equation}
The half-life $\tau$ of the decay is related to the width $\Gamma$ by well known expression:
\begin{equation}
  \tau = \hbar\; \ln 2 / \Gamma.
\label{eq.2.3.5}
\end{equation}


For description of interaction between proton and the daughter nucleus we shall use the spherical symmetric proton--nucleus potential (at case $l=0$) in Ref.~\cite{Becchetti.1969.PR}
having the following form:
\begin{equation}
  V (r, l, Q) = v_{C} (r) + v_{N} (r, Q) + v_{l} (r),
\label{eq.2.4.1}
\end{equation}
where $v_{C} (r)$, $v_{N} (r, Q)$ and $v_{l} (r)$ are Coulomb, nuclear and centrifugal components
\begin{equation}
\begin{array}{lll}
  v_{N} (r, Q) = \displaystyle\frac{V_{R}(A,Z,Q)} {1 + \exp{\displaystyle\frac{r-r_{m}} {d}}}, &
  v_{l} (r) = \displaystyle\frac{l\,(l+1)} {2mr^{2}}, &
  v_{C} (r) =
  \left\{
  \begin{array}{ll}
    \displaystyle\frac{Z e^{2}} {r}, &
      \mbox{for  } r \ge r_{m}, \\
    \displaystyle\frac{Z e^{2}} {2 r_{m}}\;
      \biggl\{ 3 -  \displaystyle\frac{r^{2}}{r_{m}^{2}} \biggr\}, &
      \mbox{for  } r < r_{m},
  \end{array}
  \right.
\end{array}
\label{eq.2.4.2}
\end{equation}
Here, $A$ and $Z$ are the nucleon and proton numbers of the daughter nucleus, $Q$ is the $Q$-value for the proton-decay, $V_{R}$ is the strength of the nuclear component, $R$ is radius of the daughter nucleus, $r_{m}$ is the effective radius of the nuclear component, $d$ is diffuseness. All parameters are defined in Ref.~\cite{Becchetti.1969.PR}.
Note that in this paper we are concentrating on the principal resolution of question to provide fully quantum basis for calculation of the penetrability and half-life in the problem of the proton decay, while the proton--nucleus potential can be used in simple form that does not take influence on the reliability of the developed methodology of multiple internal reflections absolutely and could be naturally included for modern more accurate models.


\section{Results
\label{sec.3}}


Today, there are a lot of modern methods able to calculate half-lives, which have been studied experimentally well. So, we have a rich theoretical and experimental material for analysis. We shall use these nuclei: $^{157}_{73}{\rm Ta}$, $^{161}_{75}{\rm Re}$ and $^{167}_{77}{\rm Ir}$. Such a choice we explain by that they have small coefficient of quadruple deformation $\beta_{2}$ and at good approximation can be considered as spherical (we have $l=0$). We shall study proton-decay on the basis of leaving of the particle with reduced mass from the internal region outside with its tunneling through the barrier.
This particle is supposed to start from $R_{\rm min} \le r \le R_{1}$ and move outside. Using technique of the $T^{\pm}_{j}$ and $R^{\pm}_{j}$ coefficients in eqs.~(\ref{eq.2.2.4})--(\ref{eq.2.2.6}), we calculate total
amplitudes of transmission $A_{T}$ and reflection $A_{R}$ by eqs.~(\ref{eq.2.2.8}), the penetrability coefficient $T_{MIR}$ by eqs.~(\ref{eq.2.2.9}). We check the found amplitudes, coefficients $T_{MIR}$ and $R_{MIR}$ comparing them with corresponding amplitudes and coefficients calculated by standard approach of quantum mechanics presented in Appendix~\ref{sec.app.2}.
We restrict ourselves by eq.~(\ref{eq.2.3.3}) for $F_{1}$ and find width $\Gamma$ by eq.~(\ref{eq.2.3.1}) and half-live $\tau_{MIR}$ by eq.~(\ref{eq.2.3.5}). We define the penetrability $T_{WKB}$ by
eq.~(\ref{eq.2.3.2}), calculate $\Gamma$-width and half-live $\tau_{WKB}$ by eqs.~(\ref{eq.2.3.1}) and (\ref{eq.2.3.5}).



\subsection{Dependence of the penetrability on the starting point
\label{sec.3.2}}

\begin{figure}[htbp]
\vspace{-4mm}
\centerline{\includegraphics[width=60mm]{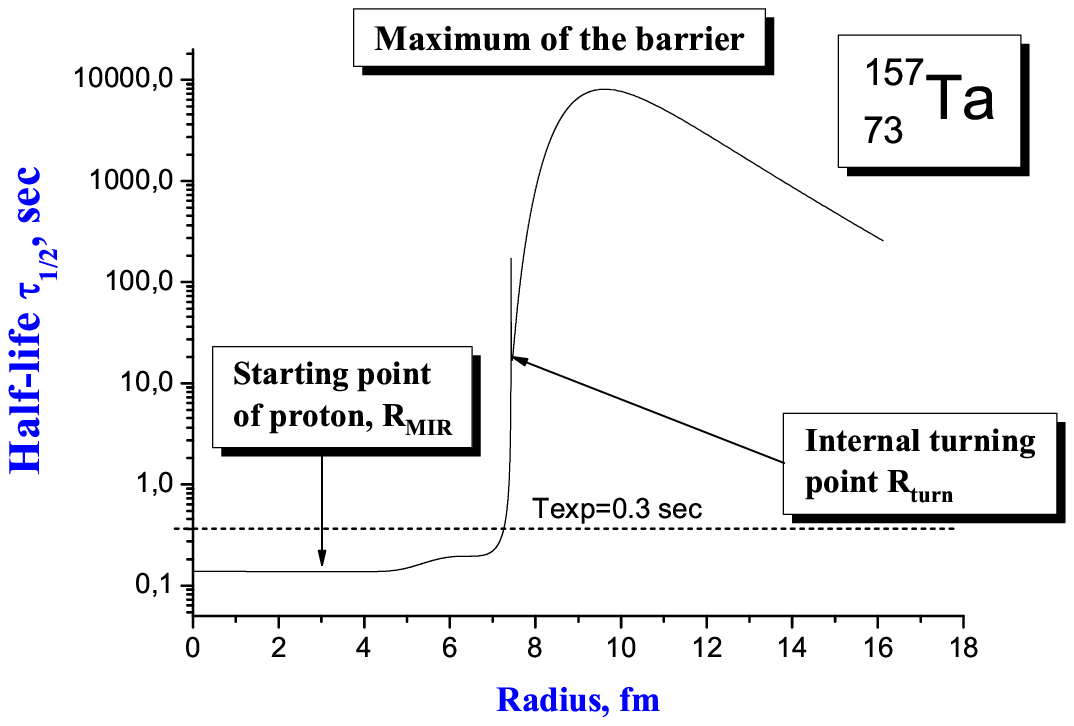}
\hspace{-4mm}\includegraphics[width=60mm]{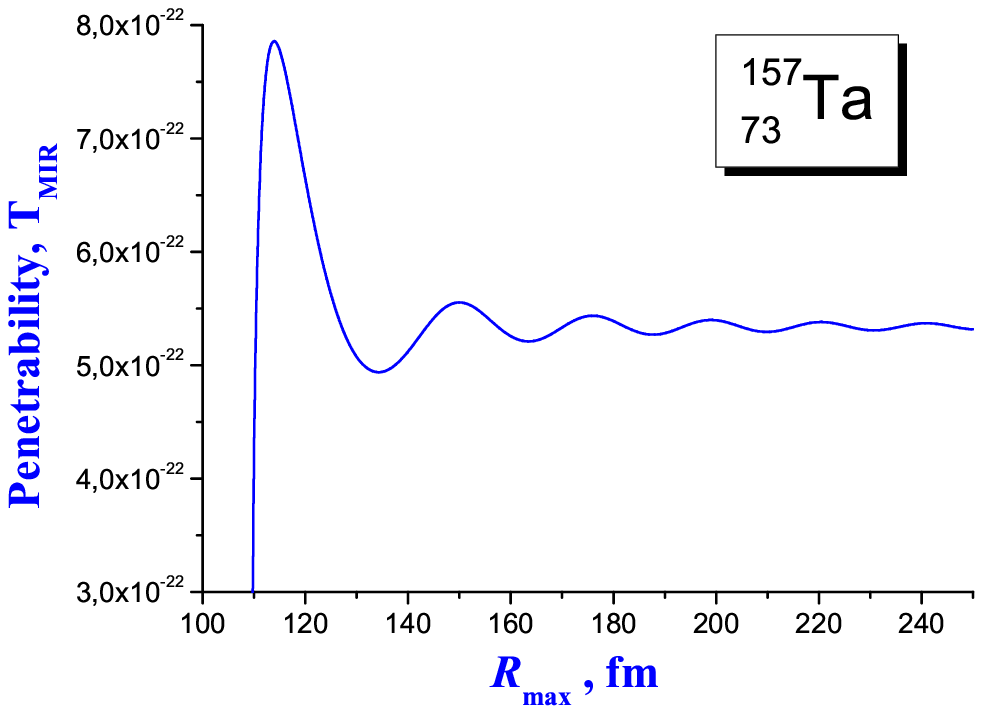}
\hspace{-4mm}\includegraphics[width=60mm]{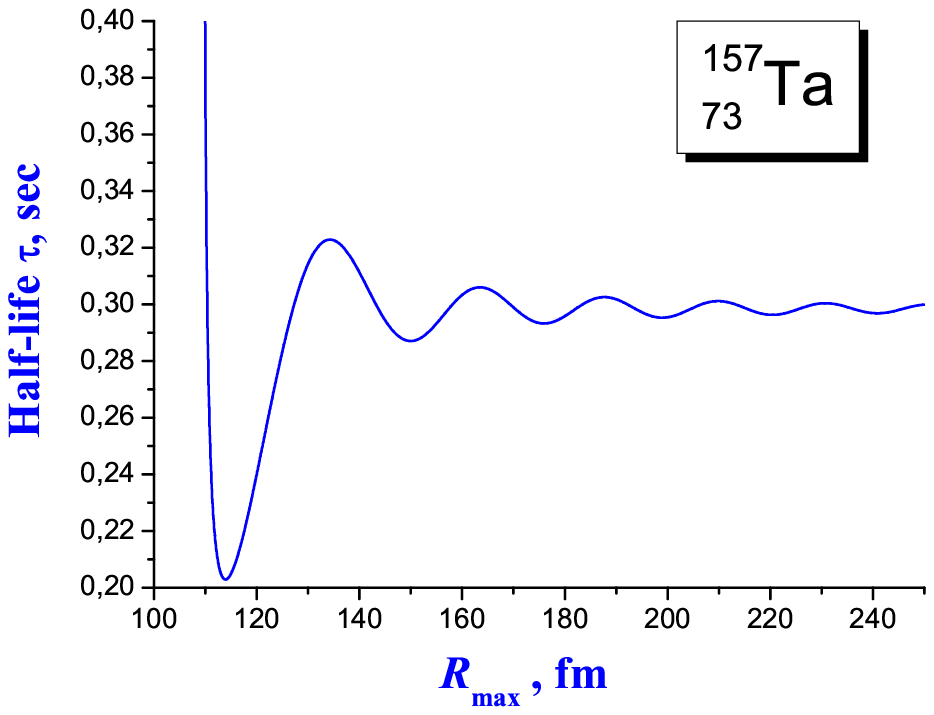}}
\vspace{-4mm} \caption{
Proton-decay for the $^{157}_{73}{\rm Ta}$ nucleus:
the left panel is for dependence of the half-life $\tau_{MIR}$ on the starting point $R_{\rm min}$,
the central panel is for dependence on penetrability $T_{MIR}$ on the external boundary $R_{\rm max}$,
the right panel is for dependence of the half-live $\tau_{MIR}$ on $R_{\rm max}$
(here, we use $R_{\rm form}=7.2127$~fm where calculated $\tau_{MIR}$ at $R_{\rm max}=250$~fm coincides with experimental data $\tau_{\rm exp}$ for this nucleus).
In all calculations factor $F$ is the same.
\label{fig.1}}
\end{figure}

The first interesting result which we have obtained is \emph{essential dependence of penetrability on the position of the first region where we localize the wave incidenting on the barrier}.
In particular, we have analyzed how much the internal boundary $R_{\rm min}$ takes influence on the penetrability. Taking into account that width of each interval is 0.01~fm, we consider point $R_{\rm min}$ as a \emph{starting point} (with error up to 0.01~fm), from here proton begins to move outside and is incident on the internal part of the barrier in the first stage of the proton decay.
In the Fig.~\ref{fig.1} [left panel] one can see that half-live of the proton decay of $^{157}_{73}{\rm Ta}$ is changed essentially at displacement of $R_{\rm min}$. So, we establish \emph{essential dependence of the penetrability on the starting point $R_{\rm start}$,
where the proton starts to move outside by approach MIR.}
At $R_{\rm form}=7.2127$~fm this dependence allows us to achieve very close coincidence between the half-live calculated by the approach MIR and experimental data.


\subsection{Dependence of the penetrability on the external region
\label{sec.3.3}}

The region of the barrier located between turning points $R_{2}$ and $R_{3}$ is main part of the potential used in calculation of the penetrability in the semiclassical approach (up to the second correction), while the internal and external parts of this potential do not take influence on it. Let us analyze whether convergence exists in calculations of the penetrability in the approach MIR if to increase the external boundary $R_{\rm max}$ ($R_{\rm max} > R_{3}$). Keeping width of each interval (step) to be the same, we shall increase $R_{\rm max}$ (through increasing number of intervals in the external region), starting from the external turning point $R_{3}$, and calculate the corresponding penetrability $T_{MIR}$. In Fig.~\ref{fig.1}~[central panel] one can see how the penetrability is changed for $^{157}_{73}{\rm Ta}$
with increasing $R_{\rm max}$.
Dependence of the half-life $\tau_{MIR}$ on $R_{\rm max}$ is shown in the next figure~\ref{fig.1}~[right panel]. One can see that the method MIR gives convergent values for the penetrability and half-life at increasing of $R_{\rm max}$. From such figures we find that \emph{inclusion of the external region into calculations changes the half-life up to 1.5 times} ($\tau_{\rm min}=0.20$~sec is the minimal half-life calculated at $R_{3} \le R_{\rm max}\le 250$~fm, $\tau_{\rm as}=0.30$~sec is the half-life calculated at $R_{\rm max}=250$~fm, ${\rm error} = \tau_{\rm as} / \tau_{\rm min} \approx 1.5$ or 50 percents). So, \emph{error in determination of the penetrability in the semiclassical approach  (if to take the external region into account) is expected to be the same as a minimum on such a basis}.



\subsection{Results for the proton emitters $^{157}_{73}{\rm Ta}$, $^{161}_{75}{\rm Re}$ and $^{167}_{77}{\rm Ir}$
\label{sec.3.4}}

So, the fully quantum study of the penetrability of the barrier for the proton decay give us its large dependence on the starting point. In order to give power of predictions of half-lives calculated by the approach MIR, we need to find recipe able to resolve such uncertainty in calculations of the half-lives. We shall introduce the following hypothesis:
\emph{we shall assume that in the first stage of the proton decay proton starts to move outside the most probably at the coordinate of minimum of the internal well}.
If such a point is located in the minimum of the well, the half-live obtains minimal value. So, as criterion we could use minimum of half-live for the given potential, which has stable basis.
Let us analyze which results such approach gives. We shall compare the half-lives calculated by approach MIR and by the semiclassical approach with experimental data. We should take into account that the half-lives obtained before are for the proton occupied ground state while it needs to take into account probability that this state is empty in the daughter nucleus. In order to obtain proper values for the half-lives we should divide them on the spectroscopic factor $S$ (which we take from \cite{Aberg.1997.PRC}), and then to compare them with experimental data.
Results of such calculations are presented in Table 1. One can see that the calculated half-lives by MIR approach turn out to be a little closer to experimental data in comparison with half-lives obtained by the semiclassical approach.

\begin{table}
\caption{Experimental and calculated half-lives of the ground state proton emitters $^{157}_{73}{\rm Ta}$, $^{161}_{75}{\rm Re}$ and $^{167}_{77}{\rm Ir}$.
Here, $S_{p}^{\rm th}$ is theoretical spectroscopic factor,
$\tau_{WKB}$ is half-life calculated by in the semiclassical approach,
$\tau_{MIR}$ is half-life calculated by in the approach MIR,
$\tilde{\tau}_{WKB} = \tau_{WKB}/\, S_{p}^{\rm th}$,
$\tilde{\tau}_{MIR} = \tau_{MIR} / S_{p}^{\rm th}$,
$\tau_{\rm exp}$ is experimental data,
$R_{\rm form}$ is starting point in the internal well where the proton begins to move outside in the first stage of the proton decay,
$R_{\rm tp}$ is turning point
(values for $S_{p}^{\rm th}$, $\tau_{\rm exp}$ are used from Table~IV in Ref.~\cite{Aberg.1997.PRC}, p.~1770;
in calculations for each nucleus we use: $R_{\rm min}=0.11$~fm, $R_{\rm max}=250$~fm;
number of intervals in region from $R_{\rm min}$ to 5~fm is 2000,
in region from 5~fm to 8~fm is 500, in region
from 8~fm to $R_{\rm max}$ is 5000)
\label{table.1}}
\begin{center}
\begin{tabular}{|c|c|c|c|c|c|c|c|c|c|} \hline
 \multicolumn{3}{|c|}{Parent nucleus} &
 \multicolumn{5}{|c|}{Half-live-values, sec} &
 \multicolumn{2}{|c|}{Points, fm} \\  \hline
  Nucleus &
  $Q$, MeV &
  $S_{p}^{\rm th}$ &
  $\tau_{WKB}$ &
  $\tau_{MIR}$ &
  $\tilde{\tau}_{WKB}$ &
  $\tilde{\tau}_{MIR}$ &
  $\tau_{\rm exp}$ &
  $R_{\rm form}$ &
  $R_{\rm 2,tp}$
  \\ \hline
    $^{157}_{73}{\rm Ta}$ & 0.947 &   
    0.66 &
    $1.313\cdot 10^{-1}$ &
    $1.369\cdot 10^{-1}$ &
    $1.99\cdot 10^{-1}$ &
    $2.074\cdot 10^{-1}$ &
    $3.00\cdot 10^{-1}$ &              
    $3.1$ &              
    7.43 \\
    $^{161}_{75}{\rm Re}$ & 1.214 &            
    0.59 &
    $1.5352\cdot 10^{-4}$ &
    $1.5314\cdot 10^{-4}$ &
    $2.602\cdot 10^{-4}$ &
    $2.596\cdot 10^{-4}$ &
    $3.70\cdot 10^{-4}$ & 
    3.32 &                
    7.34 \\               
    $^{167}_{77}{\rm Ir}$ & 1.086 &            
    0.51 &
    $2.981\cdot 10^{-2}$ &
    $2.979\cdot 10^{-2}$ &
    $5.85\cdot 10^{-2}$ &
    $5.84\cdot 10^{-2}$ &
    $1.10\cdot 10^{-2}$ & 
    3.41 &                
    7.46 \\               
%
\hline
\end{tabular}
\end{center}
\end{table}


\subsection{Comparison with other approaches of calculations of widths of proton decay
\label{sec.3.5}}

Half-life of the proton decay is defined on the basis of width $\Gamma$ which can be calculated by different approaches. For determination of width we shall use systematics of different approaches proposed
in Ref.~\cite{Aberg.1997.PRC}.
The proton emitters are narrow resonances with extremely small widths. Perturbative approach based on standard reaction theory could be expected to be accurate. Let us analyze two following approaches in such a direction.


\subsubsection{The distorted wave Born approximation method
\label{sec.3.5.1}}

The resonance width can be expressed through transition amplitude, which in the distorted wave Born approximation
(DWBA) is given so \cite{Aberg.1997.PRC}:
\begin{equation}
  T_{A+1,Z+1;\, A,Z} =
  \langle \psi_{Ap}\, \Psi_{Ap}\, \bigl| V_{Ap} \bigr|\, \Psi_{A+1} \rangle.
\label{eq.2.1.1}
\end{equation}
The DWBA calculations of the decay width requires knowledge of the quasistationary initial state wave function, $\Psi_{A+1}$, the final state wave function, $\Psi_{Ap}\,\psi_{Ap}$, and interaction potential.
The initial state wave function, $\Psi_{A+1}$, is written as a product of the daughter-nucleus wave function, $\Phi_{A}$, and the proton wave function, $\Phi_{nlj}$. The radial wave function of the proton $\psi_{l}(r) = \Psi_{l}(r)/r$ is found by numerically integrating the Schr\"{o}dinger equation with one-body potential, and it should be irregular part of the Coulomb wave function, $G_{l}(r)$, in asymptotic region. So, such wave function is complex and it defines non-zero flux. As we use condition of continuity of total flux (i.e. absence of sources inside spatial region) we cannot obtain zero wave function in whole region of its definition, and at $r=0$, in particular.

In the final state the wave function of the decaying nuclear system can be written as a product of the intrinsic wave function of the proton and the daughter nucleus (an inner core). Radial part of the proton wave function is $\psi_{l}(r) \sim F_{l}(r)/r$, where $F_{l}(r)$ is the regular Coulomb function. By other words, this wave function is real, and, therefore, it gives zero flux \underline{exactly} determined on the basis of the total wave function in the initial state.
The total wave functions in the initial and final states correspond to different processes (with different total fluxes). This confirms that they, complete wave functions, do not take reflection from the barrier inside the internal region into account (but they are defined by different boundary conditions in the initial and final states only). Here, question about determination of the decay width is passed on successful determination of perturbation of the potential (that has another \underline{physical} basis for the definition of the decay width as definition on the basis of the penetrability of the barrier). However, \emph{the question about separation of the total wave function in the internal region before the barrier into the incident and reflected waves remains unresolved in the DWBA method.}

Now, if we pass from real radial potential in optical model approach to complex one, then we shall introduce new additional independent parameter into our problem while the penetrability could be calculated for real radial barrier. Essential point in determination of the decay width in the DWBA method is accurate normalization of the wave functions in the initial and final states. It could introduce some (essential) uncertainty in calculation of width also while the penetrability is independent on such normalization absolutely.

One can calculate the decay width through time-reversed capture process. However, in such calculations shape of the barrier is approximated by inverse oscillator (or other potentials with knowing exact solutions of the wave function) and the penetrability for such a barrier could be calculated. It is clear that both internal well and external region do not take influence on results absolutely (like calculations in semiclassical approach). But, this is possible to resolve this problem accurately and taking whole studied shape of the potential barrier into account that we have demonstrated above in the fully quantum approach MIR.


\subsubsection{The two-potential method
\label{sec.3.5.2}}

In the modified two-potential approach (TPA) introduced by Gurvitz and Kalbermann in~\cite{Gurvitz.1987.PRL}
(details and examples can be found in \cite{Gurvitz.1988.PRA}, see also \cite{Jackson.1977.AP,Aberg.1997.PRC,Gurvitz.2004.PRA}) the decay width is defined so
(see (16) in \cite{Aberg.1997.PRC}, and some details):
\begin{equation}
  \Gamma =
  \displaystyle\frac{4\mu}{\hbar^{2}k}\:
    \Biggl| \displaystyle\int\limits_{r_{B}}^{\infty} \psi_{nlj}(r)\, W(r)\, \chi_{l}(r)\; dr \Biggr|^{2},
\label{eq.2.2.1}
\end{equation}
where $k=\sqrt{2\mu E_{0}} / \hbar$,
$\mu$ is reduced mass,
$r_{B}$ is radial coordinate of the barrier height,
$\psi_{nlj}(r)$ is the radial wave function for the first radial potential including internal well up to point $r_{B}$,
$\chi_{l}(r)$ is the regular radial wave function for the second radial potential including external region, starting from point $r_{B}$ and without the internal well and with asymptotic
behavior
\begin{equation}
\begin{array}{ccc}
  \chi_{l} (0) = 0, &
  \chi_{l} (r) \to \sin(kr - \pi l/2 + \delta_{l}) & {\rm at}\;  r \to \infty.
\end{array}
\label{eq.2.2.2}
\end{equation}
Both wave functions are real and defined at different energy levels. So, in the TPA approach we do not consider fluxes and do not calculate penetrability. We do not study possible reflection of proton from the barrier in the internal well. We escape from a problem of separation of the total wave function in the internal well into the incident and reflected waves which takes influence on the resulting penetrability essentially (for example, \emph{for the simplest rectangular barrier with rectangular well such an uncorrect separation of the same exact wave function can give infinite penetrability} that is explained by increased role of interference between incident and reflected waves).
Success in obtaining the resulting width $\Gamma$ is dependent on accuracy of correspondence between internal and external wave functions $\psi_{nlj}(r)$ and $\chi_{l}(r)$ which should be calculated from different Schr\"{o}dinger equations with independent normalization. The correspondence between these wave functions is determined concerning only one boundary point $r_{B}$ (or it possible shift \cite{Gurvitz.2004.PRA}) separating two potentials and boundary conditions at $r=0$ or at $r \to \infty$.
In contrary, the correspondence between the incident, transmitted and reflected waves in the MIR approach is determined concerning the barrier as the whole potential (with needed restrictions of the radial problem) that corresponds to fully quantum and unified consideration of penetration of the proton through the barrier shown in principle of \emph{non-locality} of quantum mechanics.
In particular, the transmitted wave in the MIR approach is \underline{strongly} dependent on the depth of the internal well and its shape, while the external wave function $\chi_{l}(r)$ in the TPA approach is absolutely independent on these depth and shape (such a dependence can be found in the wave function $\varphi_{nlj}(r)$, but starting from the simplest WKB approach factor F directly includes it also).
By other words, we have strong correspondence between incident, reflected and transmitted waves in the MIR approach and a possible week correspondence between the internal and external wave functions in the TPA approach.
This plays the essential role in calculations of the decay widths and explains so large difference between the essential dependence of penetrability on the starting point in the MIR approach and practically full absence of such a dependence in the TPA approach.

The simplest example demonstrated why this dependence really exists and it could be not small, can be found in classical tasks of quantum mechanics. Let us consider definition of the penetrability in~\cite{Landau.v3.1989}
(see eq.~(25.3), p.~103):
\begin{equation}
  D = \displaystyle\frac{k_{2}}{k_{1}}\: |A|^{2},
\label{eq.2.1.1}
\end{equation}
where $D$ is the penetrability, $k_{1}$ and $k_{2}$ are wave numbers of transmitted and incident waves, i.e. concerning the left asymptotic part of the potential and its asymptotic right part (see Fig.~5 in~\cite{Landau.v3.1989}, p.~103), $A$ is the transmitted amplitude of the wave function.
This formula demonstrates that decreasing of the left part of potential increases the wave number $k_{1}$ (as is connected with asymptotic presentations~(25.1) and (25.2) of waves) and, so, changes the total penetrability $D$.
Result on the essential dependence of the penetrability of the starting point $R_{\rm form}$ above has the similar sense, but has been obtained concerning the realistic barrier with the internal well and takes into account change of the internal amplitudes also. This contradicts with a possible little dependence of penetrability on the shape of the internal well in the TPA approach.
So, these points seem to be reduction of the TPA approach, and confirm that \emph{this approach does not determine the penetrability in the fully quantum consideration in the problem of proton decay}. 
At the same time, comparison of results obtained by such approach and results obtained by principally other fully quantum developments sometimes leads to some confusion as the TPA approach has been called as the fully quantum.
So, approaches for determination of the decay widths on the basis of penetrability are physically motivated, could be more accurate and have perspective for research.


\section{Conclusions
\label{conclusions}}

The new fully quantum method (called as the method of multiple internal reflections, or MIR) for calculation of widths for the decay of the nucleus by emission of proton in the spherically symmetric approximation and the realistic radial barrier of arbitrary shape is presented.
Note the following:
\begin{itemize}
\item
Solutions for amplitudes of wave function (described motion of the proton from the internal region outside with its tunneling through the barrier), penetrability $T$ and reflection $R$ are found by the method MIR for $n$-step radial barrier at arbitrary $n$. These solutions are \emph{exactly solvable} and have been obtained in \emph{the fully quantum approach} for the first time. At limit $n \to \infty$ these solutions could be considered as exact ones for the realistic proton--nucleus potential with needed arbitrary barrier and internal hole. Estimated error of the achieved results is $|T+R-1| < 1.5 \cdot 10^{-15}$.

\item
In contrast to the semiclassical approach and the TPA approach, the approach MIR gives essential dependence of the penetrability on the starting point $R_{\rm form}$ inside the internal well where proton starts to move outside in the beginning of the proton decay. For example, the penetrability of the barrier calculated by MIR approach for $^{157}_{73}{\rm Ta}$ is changed up to 200 times in dependence on position of $R_{\rm form}$ (see Fig.~\ref{fig.1}, the left panel).
The amplitudes calculated by MIR approach we compared with the corresponding amplitudes obtained (for the same potential) by independent standard stationary approach of quantum mechanics presented in Appendix~\ref{sec.app.2} and we obtained coincidence up to first 15 digits for all considered amplitudes.
This important test confirms that \emph{presence of the essential dependence of the penetrability of the starting point $R_{\rm form}$ is result independent on the fully quantum method applied}.
Such a result could be connected with a possibility to introduce initial condition which could be imposed on proton decay in the fully quantum consideration.
Comparison with the WKB and TPA approaches shows that such approaches have no such a perspective (having physical sense and opening a possibility to obtain a new information about the proton decay), which fully quantum study of the penetrability gives.

\item
In order to resolve uncertainty in calculations of the half-lives caused by the dependence of the penetrability on $R_{\rm form}$, we have introduced the hypothesis:
\emph{in the first stage of the proton decay the proton starts to move outside at the coordinate of minimum of the internal well}. Such condition provides minimal value for the calculated half-life and gives stable basis for predictions in the MIR approach. However, the half-lives calculated by the MIR approach turn out to be a little closer to experimental data in comparison with the half-lives obtained by the semiclassical approach (see Tabl.~1).

\item
Taking the external region of the potential after the barrier into account, half-live calculated by the MIR approach is changed up to 1.5 times (see Fig.~\ref{fig.1}, the right panel).
\end{itemize}
A main advance of the MIR method developed in this paper is not a new attempt to describe experimental data of half-lives more accurately than other approaches do this, but rather this method seems to be the first tools for estimation of the penetrability of any desirable barrier of the proton decay in the fully quantum consideration.


\appendix
\section{Tunneling of packet through one-dimensional rectangular step
\label{sec.app.1}}

\begin{figure}[htbp]
\centerline{\includegraphics[width=50mm]{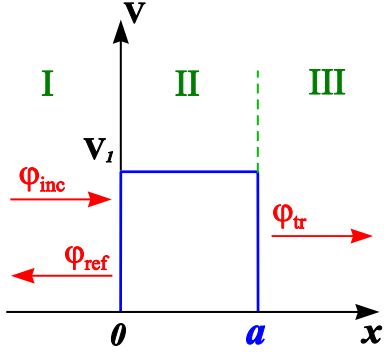}}
\caption{\small
Tunneling of the particle through one-dimensional rectangular barrier
\label{fig.2}}
\end{figure}

Main ideas and formalism of the multiple internal reflections can be the most clearly analyzed in the simplest problem of tunneling of the particle through one-dimensional rectangular barrier in whole axis
\cite{Maydanyuk.2000.UPJ,Maydanyuk.2002.JPS,Maydanyuk.2002.PAST,Maydanyuk.2003.PhD-thesis,Maydanyuk.2006.FPL}.
Let us consider a problem of tunneling of a particle in a positive $x$-direction through an one-dimensional rectangular potential barrier (see Fig.~\ref{fig.2}). Let us label a region I for $x < 0$, a region II for $0 < x < a$ and a region III for $x > a$, accordingly.
We shall study an evolution of its tunneling through the barrier.
In standard approach, with energy less than the barrier height the tunneling evolution of the particle is described using a non-stationary propagation of WP
\begin{equation}
  \psi(x, t) = \int\limits_{0}^{+\infty} g(E - \bar{E}) \varphi(k, x)
                e^{-iEt/\hbar} dE,
\label{eq.2.1.1}
\end{equation}
where stationary WF is:
\begin{equation}
\varphi(x) = \left\{
\begin{array}{ll}
   e^{ikx}+A_{R}e^{-ikx},               & \mbox{for } x<0,   \\
   \alpha e^{\xi x} + \beta e^{-\xi x}, & \mbox{for } 0<x<a, \\
   A_{T} e^{ikx},                       & \mbox{for } x>a
\end{array} \right.
\label{eq.2.1.2}
\end{equation}
and $k   = \frac{1}{\hbar}\sqrt{2mE}$, $\xi = \frac{1}{\hbar}\sqrt{2m(V_{1}-E)}$,
$E$ and $m$ are the total energy and mass of the particle, accordingly.
The weight amplitude $g(E - \bar{E})$ can be written in the standard gaussian form and satisfies to a requirement of the normalization $\int |g(E - \bar{E})|^{2} dE = 1$, value $\bar{E}$ is an average energy of the particle. One can calculate coefficients $A_{T}$, $A_{R}$, $\alpha$ and $\beta$ analytically, using a requirements of a continuity of WF $\varphi(x)$ and its derivative on each boundary of the barrier.
Substituting in eq.~(\ref{eq.2.1.1}) instead of $\varphi(k, x)$ the incident $\varphi_{inc}(k, x)$, transmitted  $\varphi_{tr}(k, x)$ or reflected part of WF $\varphi_{ref}(k, x)$, defined by eq.~(\ref{eq.2.1.2}), we receive the incident, transmitted  or reflected WP, accordingly.

We assume, that a time, for which the WP tunnels through the barrier, is enough small. So, the time necessary for a tunneling of an $\alpha$-particle through a barrier of decay in $\alpha$-decay of a nucleus, is about $10^{-21}$ seconds. We consider, that one can neglect a spreading of the WP for this time. And a breadth of the WP appears essentially more narrow on a comparison with a barrier breadth.
Considering only sub-barrier processes, we exclude a component of waves for above-barrier energies, having included the additional transformation
\begin{equation}
   g(E - \bar{E}) \to g(E - \bar{E}) \theta(V_{1} - E),
\label{eq.2.1.3}
\end{equation}
where $\theta$-function satisfies to the requirement
%
\[
\theta(\eta) = \left\{
  \begin{array}{ll}
    0, & \mbox{for } \eta<0;   \\
    1, & \mbox{for } \eta>0.
  \end{array} \right.
\]

The method of multiple internal reflections considers the propagation process of the WP describing a motion of the particle, sequentially on steps of its penetration in relation to each boundary of the barrier
\cite{Fermor.1966.AJPIA,McVoy.1967.RMPHA,Anderson.1989.AJPIA}. Using this method, we find expressions for the transmitted and reflected WP in relation to the barrier.
At the first step we consider the WP in the region I, which is incident upon the first (initial) boundary of the barrier. Let's assume, that this package transforms into the WP, transmitted through this boundary and tunneling further in the region II, and into the WP, reflected from the boundary and propagating back in the region I. Thus we consider, that the WP, tunneling in the region II, is not reached the second (final) boundary of the barrier because of a terminating velocity of its propagation, and consequently at this step we consider only two regions I and II. Because of physical reasons to construct an expression for this packet, we consider, that its amplitude should decrease in a positive $x$-direction. We use only one item $\beta\exp(-\xi x)$ in eq.~(\ref{eq.2.1.2}), throwing the second increasing item $\alpha\exp(\xi x)$ (in an opposite case we break a requirement of a finiteness of the WF for an indefinitely wide barrier).
In result, in the region II we obtain:
\begin{equation}
  \psi^{1}_{tr}(x, t) = \int\limits_{0}^{+\infty} g(E - \bar{E})
  \theta(V_{1} - E) \beta^{0} e^{-\xi x -iEt/\hbar} dE,
  \mbox{for } 0<x<a.
\label{eq.2.1.4}
\end{equation}
Thus the WF in the barrier region constructed by such way, is an analytic continuation of a relevant expression for the WF, corresponding to a similar problem with above-barrier energies, where as a stationary expression we select the wave $\exp(ik_{2}x)$, propagated to the right.

Let's consider the first step further. One can write expressions for the incident and the reflected WP in relation to the first boundary as follows
\begin{equation}
\begin{array}{lcll}
\psi_{inc}(x, t) & = & \int\limits_{0}^{+\infty} g(E - \bar{E})
        \theta(V_{1} - E) e^{ikx -iEt/\hbar} dE,
        & \mbox{for } x<0, \\
\psi^{1}_{ref}(x, t) & = & \int\limits_{0}^{+\infty} g(E - \bar{E})
        \theta(V_{1} - E) A_{R}^{0} e^{-ikx -iEt/\hbar} dE,
        & \mbox{for } x<0.
\end{array}
\label{eq.2.1.5}
\end{equation}
A sum of these expressions represents the complete WF in the region I, which is dependent on a time. Let's require, that this WF and its derivative continuously transform into the WF (\ref{eq.2.1.4}) and its derivative at point $x=0$
(we assume, that the weight amplitude $g(E - \bar{E})$ differs weakly at transmitting and reflecting of the WP in relation to the barrier boundaries). In result, we obtain two equations, in which one can pass from the time-dependent WP to the corresponding stationary WF and obtain the unknown coefficients $\beta^{0}$ and $A_{R}^{0}$.

At the second step we consider the WP, tunneling in the region II and incident upon the second boundary of the barrier at point $x = a$. It transforms into the WP, transmitted through this boundary and propagated in the region III, and into the WP, reflected from the boundary and tunneled back in the region II. For a determination of these packets one can use eq.~(\ref{eq.2.1.1}) with account eq.~(\ref{eq.2.1.3}),
where as the stationary WF we use:
\begin{equation}
\begin{array}{lcll}
\varphi_{inc}^{2}(k, x) & = & \varphi_{tr}^{1}(k, x) =
        \beta^{0} e^{-\xi x},
        & \mbox{for } 0<x<a, \\
\varphi_{tr}^{2}(k, x) & = & A_{T}^{0}e^{ikx},
        & \mbox{for } x>a, \\
\varphi_{ref}^{2}(k, x) & = & \alpha^{0} e^{\xi x},
        & \mbox{for } 0<x<a.
\end{array}
\label{eq.2.1.6}
\end{equation}
Here, for forming an expression for the WP reflected from the boundary, we select an increasing part of the stationary solution $\alpha^{0} \exp(\xi x)$ only. Imposing a condition of continuity on the time-dependent WF and its derivative at point $x = a$, we obtain 2 new equations, from which we find the unknowns coefficients $A_{T}^{0}$ and $\alpha^{0}$.

At the third step the WP, tunneling in the region II, is incident upon the first boundary of the barrier. Then it transforms into the WP, transmitted through this boundary and propagated further in the region I, and into the WP, reflected from boundary and tunneled back in the region II. For a determination of these packets one can use eq.~ (\ref{eq.2.1.1}) with account eq.~(\ref{eq.2.1.3}), where as the stationary WF we use:
\begin{equation}
\begin{array}{lcll}
\varphi_{inc}^{3}(k, x) & = & \varphi_{ref}^{2}(k, x),
        & \mbox{for } 0<x<a, \\
\varphi_{tr}^{3}(k, x) & = & A_{R}^{1}e^{-ikx},
        & \mbox{for } x<0, \\
\varphi_{ref}^{3}(k, x) & = & \beta^{1} e^{-\xi x},
        & \mbox{for } 0<x<a.
\end{array}
\label{eq.2.1.7}
\end{equation}
Using a conditions of continuity for the time-dependent WF and its derivative at point $x = 0$, we obtain the unknowns coefficients $A_{R}^{1}$ and $\beta^{1}$.

Analyzing further possible processes of the transmission (and the reflection) of the WP through the boundaries of the barrier, we come to a deduction, that any of following steps can be reduced to one of 2 considered above. For the unknown coefficients $\alpha^{n}$, $\beta^{n}$,$A_{T}^{n}$ and $A_{R}^{n}$, used in expressions for the WP, forming in result of some internal reflections from the boundaries, one can obtain the recurrence relations:
\begin{equation}
\begin{array}{lll}
\beta^{0} = \displaystyle\frac{2k}{k+i\xi},     &
\alpha^{n} = \beta^{n} \displaystyle\frac{i\xi-k}{i\xi+k}e^{-2\xi a}, &
\beta^{n+1} = \alpha^{n} \displaystyle\frac{i\xi-k}{i\xi+k}, \\
A_{R}^{0} = \displaystyle\frac{k-i\xi}{k+i\xi},     &
A_{T}^{n} = \beta^{n} \displaystyle\frac{2i\xi}{i\xi+k}e^{-\xi a-ika}, &
A_{R}^{n+1} = \alpha^{n} \displaystyle\frac{2i\xi}{i\xi+k}.
\end{array}
\label{eq.2.1.8}
\end{equation}

Considering the propagation of the WP by such way, we obtain expressions for the WF on each region which can be written through series of multiple WP. Using eq.~(\ref{eq.2.1.1}) with account eq.~(\ref{eq.2.1.3}), we determine resultant expressions for the incident, transmitted and reflected WP in relation to the barrier, where one can need to use following expressions for the stationary WF:
\begin{equation}
\begin{array}{lcll}
\varphi_{inc}(k, x) & = & e^{ikx},
                        & \mbox{for } x<0, \\
\varphi_{tr}(k, x)  & = & \sum\limits_{n=0}^{+\infty} A_{T}^{n} e^{ikx},
                        & \mbox{for } x>a, \\
\varphi_{ref}(k, x) & = & \sum\limits_{n=0}^{+\infty} A_{R}^{n} e^{-ikx},
                        & \mbox{for } x<0.
\end{array}
\label{eq.2.1.9}
\end{equation}

Now we consider the WP formed in result of sequential $n$ reflections from the boundaries of the barrier and incident upon one of these boundaries at point $x = 0$ ($i = 1$) or at point $x = a$ ($i = 2$). In result, this WP transforms into the WP $\psi_{tr}^{i}(x, t)$, transmitted through boundary with number $i$, and into the WP $\psi_{ref}^{i}(x, t)$, reflected from this boundary. For an independent on $x$ parts of the stationary WF one can write:
\begin{equation}
\begin{array}{ll}
   \displaystyle\frac{\varphi_{tr}^{1}}{\exp(-\xi x)} =
   T_{1}^{+} \displaystyle\frac{\varphi_{inc}^{1}}{\exp(ikx)}, &
   \displaystyle\frac{\varphi_{ref}^{1}}{\exp(-ikx)} =
   R_{1}^{+} \displaystyle\frac{\varphi_{inc}^{1}}{\exp(ikx)}, \\
   \displaystyle\frac{\varphi_{tr}^{2}}{\exp(ikx)} =
   T_{2}^{+} \displaystyle\frac{\varphi_{inc}^{2}}{\exp(-\xi x)}, &
   \displaystyle\frac{\varphi_{ref}^{2}}{\exp(\xi x)} =
   R_{2}^{+} \displaystyle\frac{\varphi_{inc}^{2}}{\exp(-\xi x)}, \\
   \displaystyle\frac{\varphi_{tr}^{1}}{\exp(-ikx)} =
   T_{1}^{-} \displaystyle\frac{\varphi_{inc}^{1}}{\exp(\xi x)}, &
   \displaystyle\frac{\varphi_{ref}^{1}}{\exp(-\xi x)} =
   R_{1}^{-} \displaystyle\frac{\varphi_{inc}^{1}}{\exp(\xi x)},
\end{array}
\label{eq.2.1.10}
\end{equation}
where the sign ``+'' (or ``-'') corresponds to the WP, tunneling (or propagating) in a positive (or negative) $x$-direction and incident upon the boundary with number $i$. Using $T_{i}^{\pm}$ and $R_{i}^{\pm}$, one can precisely describe an arbitrary WP which has formed in result of $n$-multiple reflections, if to know a ``path'' of its propagation along the barrier. Using the recurrence relations eq.~(\ref{eq.2.1.8}), the coefficients $T_{i}^{\pm}$ and $R_{i}^{\pm}$ can be obtained.
\begin{equation}
\begin{array}{lll}
T_{1}^{+} = \beta^{0},
&
T_{2}^{+} = \displaystyle\frac{A_{T}^{n}}{\beta^{n}},
&
T_{1}^{-} = \displaystyle\frac{A_{R}^{n+1}}{\alpha^{n}},
\\
R_{1}^{+} = A_{R}^{0},
&
R_{2}^{+} = \displaystyle\frac{\alpha^{n}}{\beta^{n}},
&
R_{1}^{-} = \displaystyle\frac{\beta^{n+1}}{\alpha^{n}}.
\end{array}
\label{eq.2.1.11}
\end{equation}

Using the recurrence relations, one can find series of coefficients $\alpha^{n}$, $\beta^{n}$, $A_{T}^{n}$ and $A_{R}^{n}$. However, these series can be calculated easier, using coefficients $T_{i}^{\pm}$ and $R_{i}^{\pm}$. Analyzing all possible ``paths'' of the WP propagations along the barrier, we receive:
\begin{equation}
\begin{array}{lcl}
\sum\limits_{n=0}^{+\infty} A_{T}^{n} & = &
        T_{2}^{+}T_{1}^{-} \biggl(1 + \sum\limits_{n=1}^{+\infty}
                                 (R_{2}^{+}R_{1}^{-})^{n} \biggr) =
        \displaystyle\frac{i4k \xi e^{-\xi a-ika}}{F_{sub}},  \\
\sum\limits_{n=0}^{+\infty} A_{R}^{n} & = &
        R_{1}^{+} + T_{1}^{+}R_{2}^{+}T_{1}^{-} \biggl(1 +
        \sum\limits_{n=1}^{+\infty}(R_{2}^{+}R_{1}^{-})^{n} \biggr) =
        \displaystyle\frac{k_{0}^{2}D_{-}}{F_{sub}},         \\
\sum\limits_{n=0}^{+\infty} \alpha^{n} & = &
        \alpha^{0} \biggl(1 + \sum\limits_{n=1}^{+\infty}
                           (R_{2}^{+}R_{1}^{-})^{n} \biggr) =
        \displaystyle\frac{2k(i\xi - k)e^{-2\xi a}}{F_{sub}}, \\
\sum\limits_{n=0}^{+\infty} \beta^{n} & = &
        \beta^{0} \biggl(1 + \sum\limits_{i=1}^{+\infty}
                          (R_{2}^{+}R_{1}^{-})^{n} \biggr) =
        \displaystyle\frac{2k(i\xi + k)}{F_{sub}}, \\
\end{array}
\label{eq.2.1.12}
\end{equation}
where
\begin{equation}
\begin{array}{lll}
F_{sub} & = & (k^{2} - \xi^{2})D_{-} + 2ik\xi D_{+},            \\
D_{\pm}   & = & 1 \pm e^{-2\xi a},                              \\
k_{0}^{2} & = & k^{2} + \xi^{2} = \displaystyle\frac{2mV_{1}}{\hbar^{2}}.
\end{array}
\label{eq.2.1.13}
\end{equation}

All series $\sum \alpha^{n}$, $\sum \beta^{n}$, $\sum A_{T}^{n}$ and $\sum A_{R}^{n}$, obtained using the method of multiple internal reflections, coincide with the corresponding coefficients $\alpha$, $\beta$, $A_{T}$ and $A_{R}$ of the eq.~(\ref{eq.2.1.2}), calculated by a stationary methods \cite{Landau.v3.1989}.
Using the following substitution
\begin{equation}
  i\xi \to k_{2},
\label{eq.2.1.14}
\end{equation}
where $k_{2}= \frac{1}{\hbar}\sqrt{2m (E-V_{1})}$ is a wave number for a case of above-barrier energies, expression for the coefficients $\alpha^{n}$, $\beta^{n}$, $A_{T}^{n}$ and $A_{R}^{n}$ for each step, expressions for the WF for each step, the total eqs.~(\ref{eq.2.1.12}) and (\ref{eq.2.1.13}) transform into the corresponding expressions for a problem of the particle propagation above this barrier. At the transformation of the WP and the time-dependent WF one can need to change a sign of argument at $\theta$-function. Besides the following property is fulfilled:
\begin{equation}
  \biggl|\sum\limits_{n=0}^{+\infty} A_{T}^{n}\biggr|^{2} +
  \biggl|\sum\limits_{n=0}^{+\infty} A_{R}^{n}\biggr|^{2} = 1.
\label{eq.2.1.15}
\end{equation}
\section{Direct method
\label{sec.app.2}}

We shall add shortly solution for amplitudes of the wave function obtained by standard technique of quantum mechanics which could be obtained if to use only condition of continuity of the wave function and its derivative at each boundary, but on the whole region of the studied potential. At first, we find functions $f_{2}$ and $g_{2}$
(from the first boundary):
\begin{equation}
\begin{array}{cc}
  f_{2} = \displaystyle\frac{k_{2}+k}{k_{2}-k} \,e^{2ik_{2}x_{1}}, &
  g_{2} = \displaystyle\frac{2k}{k-k_{2}} \,e^{i(k+k_{2})x_{1}}.
\end{array}
\label{eq.2.3.2.1}
\end{equation}
Then, using the following recurrent relations:
\begin{equation}
\begin{array}{ccl}
  f_{j+1} & = & \displaystyle\frac
              {(k_{j+1}-k_{j})\, e^{2ik_{j}x_{j}} + f_{j}\, (k_{j+1}+k_{j}) }
              {(k_{j+1}+k_{j})\, e^{2ik_{j}x_{j}} + f_{j}\, (k_{j+1}-k_{j}) }
              \cdot e^{2ik_{j+1}x_{j}},
\end{array}
\label{eq.2.3.2.2}
\end{equation}
we calculate next functions $f_{3}$, $f_{4}$, $f_{5}$ \ldots $f_{n}$,
and by such a formula:
\begin{equation}
\begin{array}{ccl}
  g_{j+1} & = & g_{j} \cdot \displaystyle\frac{2k_{j}\, e^{i(k_{j+1}+k_{j})x_{j}}}
              {(k_{j+1}+k_{j})\, e^{2ik_{j}x_{j}} + f_{j}\, (k_{j+1}-k_{j})}
\end{array}
\label{eq.2.3.2.3}
\end{equation}
the functions $g_{3}$, $g_{4}$, $g_{5}$ \ldots $g_{n}$.
From $f_{n}$ and $g_{n}$ we find amplitudes $\alpha_{n}$, $\beta_{n}$ and amplitude of transmission $A_{T}$:
\begin{equation}
\begin{array}{cc}
  \beta_{n} = 0, &
  A_{T} = \alpha_{n} = -\displaystyle\frac{g_{n}} {f_{n}}.
\end{array}
\label{eq.2.3.2.4}
\end{equation}
Now using the recurrent relations:
\begin{equation}
\begin{array}{ccl}
  \alpha_{j-1} & = & \displaystyle\frac
          {\alpha_{j}\, e^{ik_{j}x_{j-1}} + \beta_{j}\, e^{-ik_{j}x_{j-1}} - g_{j-1}\, e^{-ik_{j-1}x_{j-1}}}
          {e^{ik_{j-1}x_{j-1}} + f_{j-1}\, e^{-ik_{j-1}x_{j-1}}}
\end{array}
\label{eq.2.3.2.5}
\end{equation}
and such a formula:
\begin{equation}
  \beta_{j} = \alpha_{j} \cdot f_{j} + g_{j},
\label{eq.2.3.2.6}
\end{equation}
we consistently calculate the amplitudes $\alpha_{n-1}$, $\beta_{n-1}$, $\alpha_{n-2}$, $\beta_{n-2}$ \ldots $\alpha_{2}$, $\beta_{2}$.
At finishing, we find amplitude of reflection $A_{R}$:
\begin{equation}
  A_{R} = \alpha_{2}\,e^{i(k+k_{2})x_{1}} + \beta_{2}\,e^{i(k-k_{2})x_{1}} - e^{2 ikx_{1}}.
\label{eq.2.3.2.7}
\end{equation}
As test we use condition:
\begin{equation}
  \displaystyle\frac{k_{n}}{k_{1}}\; |A_{T}|^{2} + |A_{R}|^{2} = 1.
\label{eq.2.3.2.8}
\end{equation}
Studying the problem of proton decay, we used such a techniques for check the amplitudes obtained previously by the MIR approach and obtained coincidence up to first 15 digits for all considered amplitudes. In particular, we reconstruct completely the pictures of the probability presented in Fig.~\ref{fig.1}~(a) and (b), but using standard technique above. So, \emph{result on the large dependence of the penetrability of the position of the starting point $R_{\rm form}$ in such figures is independent on the used method}.



\end{document}